\documentclass[fleqn,10pt]{wlscirep}

\usepackage[utf8]{inputenc}
\usepackage[T1]{fontenc}
\usepackage{txfonts}
\usepackage{threeparttable}
\usepackage{amssymb}
\usepackage{amsmath}
\usepackage{kantlipsum}
\usepackage{multicol}
\usepackage[printwatermark]{xwatermark}
\usepackage{graphicx}
\usepackage{array}
\usepackage{color}
\usepackage{url}
\usepackage{enumitem}
\usepackage{bm}
\usepackage{xcolor}
\usepackage{afterpage}

\newcommand{\Rp}{R_{p}}

\newcommand{\Mp}{M_{p}}
\newcommand{\Mb}{M_{b}}
\newcommand{\Ms}{M_{*}}
\newcommand{\Msun}{M_{\odot}}
\newcommand{\Mj}{\textrm{M}_\mathrm{J}}

\newcommand{\be}{\begin{equation}}
\newcommand{\ee}{\end{equation}}

\title{Listening to the gravitational wave sound of circumbinary exoplanets}

\author[1,*]{Nicola Tamanini}
\author[2,3,*]{Camilla Danielski}
\affil[1]{Max-Planck-Institut f\"ur Gravitationsphysik, Albert-Einstein-Institut, Am M\"uhlenberg 1,14476 Potsdam-Golm, Germany. email : nicola.tamanini@aei.mpg.de}
\affil[2]{AIM, CEA, CNRS, Universit\'e Paris-Saclay, Universit\'e Paris Diderot, Sorbonne Paris Cit\'e, F-91191 Gif-sur-Yvette, France. 
email : camilla.danielski@cea.fr}
\affil[3]{Institut d'Astrophysique de Paris, CNRS, UMR 7095, Sorbonne Universit\'e, 98 bis bd Arago, 75014 Paris, France}
\affil[*]{\textit{These authors contributed equally to this work.}}

\begin{abstract} 
\textbf{
To date more than 3500 exoplanets have been discovered orbiting a large variety of stars.
Due to the sensitivity limits of the currently used detection techniques, these planets populate zones restricted either to the solar neighbourhood or towards the Galactic bulge. This selection problem prevents us from unveiling the true Galactic planetary population and is not set to change for the next two decades.
Here we present a new detection method that overcomes this issue and that will allow us to detect gas giant exoplanets using gravitational wave astronomy. We show that the Laser Interferometer Space Antenna (LISA) mission can characterise hundreds of new circumbinary exoplanets orbiting white dwarf binaries everywhere in our Galaxy -- a population of exoplanets so far completely unprobed -- as well as detecting extragalactic bound exoplanets in the Magellanic Clouds. Such a method is not limited by stellar activity and, in extremely favourable cases, will allow LISA to detect super-Earths down to 30 Earth masses.
}
\end{abstract}
\smallskip

\begin{document}

\flushbottom
\maketitle

\thispagestyle{empty}

\section{Exoplanets beyond the solar neighbourhood}

In the last twenty years the field of extrasolar planets has witnessed an exceptionally fast development, revealing an incredibly diverse menagerie of planetary companions.
These discoveries changed the place that our Solar System occupies in the Galactic context and allowed us to develop a deeper understanding of the planetary population around us.
For instance we now know that hot-Jupiters are rare, super-Earths are ubiquitous\cite{Winn2018}, and that there is a gap in the radius distribution of small planets\cite{Fulton2017}.
Nonetheless, our knowledge is restricted to the solar neighbourhood because the most successful detection techniques, such as radial velocity and transit, can only observe bright stars close to us. Differently, gravitational microlensing can observe farther away towards the Galactic bulge, but it does not provide enough targets to develop a robust statistics of the bulge population.
Determining if what we see it is given by a selection bias or not, it is extremely important, and it cannot be assessed through the usual techniques for at least the next two decades. 

This breakthrough is going to be possible only through gravitational wave (GW) astronomy. The Laser Interferometer Space Antenna (LISA) mission\cite{LISAcallpaper}, planned for launch in the early 2030s, will enable us to indirectly probe for the first time the population of gas giants orbiting detached double white dwarfs (DWD) binaries, everywhere in the Milky Way and in the nearby Magellanic Clouds. 
Given that about half of the stellar population resides in multiple stellar systems\cite{Raghavan2010,DucheneKraus2013}, and that $\sim 95 \%$ of the stars will become a white dwarf\cite{Althaus2010}, the LISA circumbinary planets survey will shed light on the final fate of an exoplanet as well as provide a galactic statistic of these objects.\\
Among the 90+ circumbinary systems currently known in the solar proximity, two tens are P-type, meaning systems with planet(s) orbiting $both$ stars in the binary. Of these, only 6 systems have a white dwarf as binary component. Usually the second companion is a low-mass star\cite{Veras2016} or, for one specific case, a pulsar \cite{Sigurdsson2003}. Today, no exoplanets have been discovered 
around double white dwarfs, irrespectively of the compactness of the binary.

Due to the intrinsic faintness of these DWDs, 
only few tens are known by spectroscopic and variability surveys\cite{Korol2017}, but substantial progress in the detection of these sources is expected via GWs.
The LISA mission, working in the gravitational wave low frequency range between 0.1 mHz and 1 Hz, will detect around 25x10$^{3}$ compact DWDs within and outside the Milky Way\cite{Korol2017,Korol2018},  some of which could be perturbed by the presence of a third gravitationally bound stellar companion\cite{Robson2018} or planetary companion.

The existence of this planetary population in our Galaxy is far from being excluded\cite{Veras2016, Kostov2016},  yet there is no observational proof of its presence.
We show here that LISA represents an important step forward in the science of planetary formation and evolution.
In case of positive planetary detections, LISA will provide crucial elements $(i)$  to understand under which conditions a planet can survive the most critical phases of a close binary evolution\cite{Zuckerman2010, Kostov2016}; $(ii)$  to set constraints on binary mass-loss and the dynamical aspects which directly follow\cite{Kostov2016,VerasTout2012}; $(iii)$ to check whether a second-generation of planets exists, i.e., planets that form from the stellar material ejected during the binary common envelope phase(s)\cite{Schleicher2014}.
Conversely, in case of no detection all over the Milky Way, LISA will enable us to set unbiased constraints on planetary evolution theories, and in particular on the fate of exoplanets bound to a binary that undergoes two common envelope phases \cite{Kostov2016}.

In what follows we assess the potential of LISA to discover new P-type circumbinary exoplanets through their perturbation on the GW signal emitted by the DWD.
This method is fully original and relies on the large DWD population of GWs sources to be heard by LISA, which makes it more powerful and interesting than former ideas of direct detection of exoplanets through GWs\cite{Ferrari2000,Berti:2000jv,Ain:2015mea,Cunha:2018ttg,Wong:2018amf}.
Finally, in the spirit of the new era of multi-messenger astronomy, we discuss the possibilities that could open for the field of exoplanets when standard electromagnetic (EM) techniques will work in synergy with GW astronomy.


\section{Detecting exoplanets with LISA}
\label{sec:results}

Galactic DWDs with periods shorter than one hour emit almost monochromatic GWs in the LISA frequency band.
If a third companion object orbits the DWD then the centre of mass (CoM) of the DWD is on a Keplerian orbit.
Through the Doppler effect, the motion of the DWD CoM induces an observable imprint on the GW waveform measured by LISA\cite{Robson2018}.
We refer to the Methods section for the dynamical model employed to describe the three-body system, and details regarding the Fisher matrix parameter estimation with LISA.

\begin{figure}[t]
\centering
\includegraphics[width=0.8\textwidth]{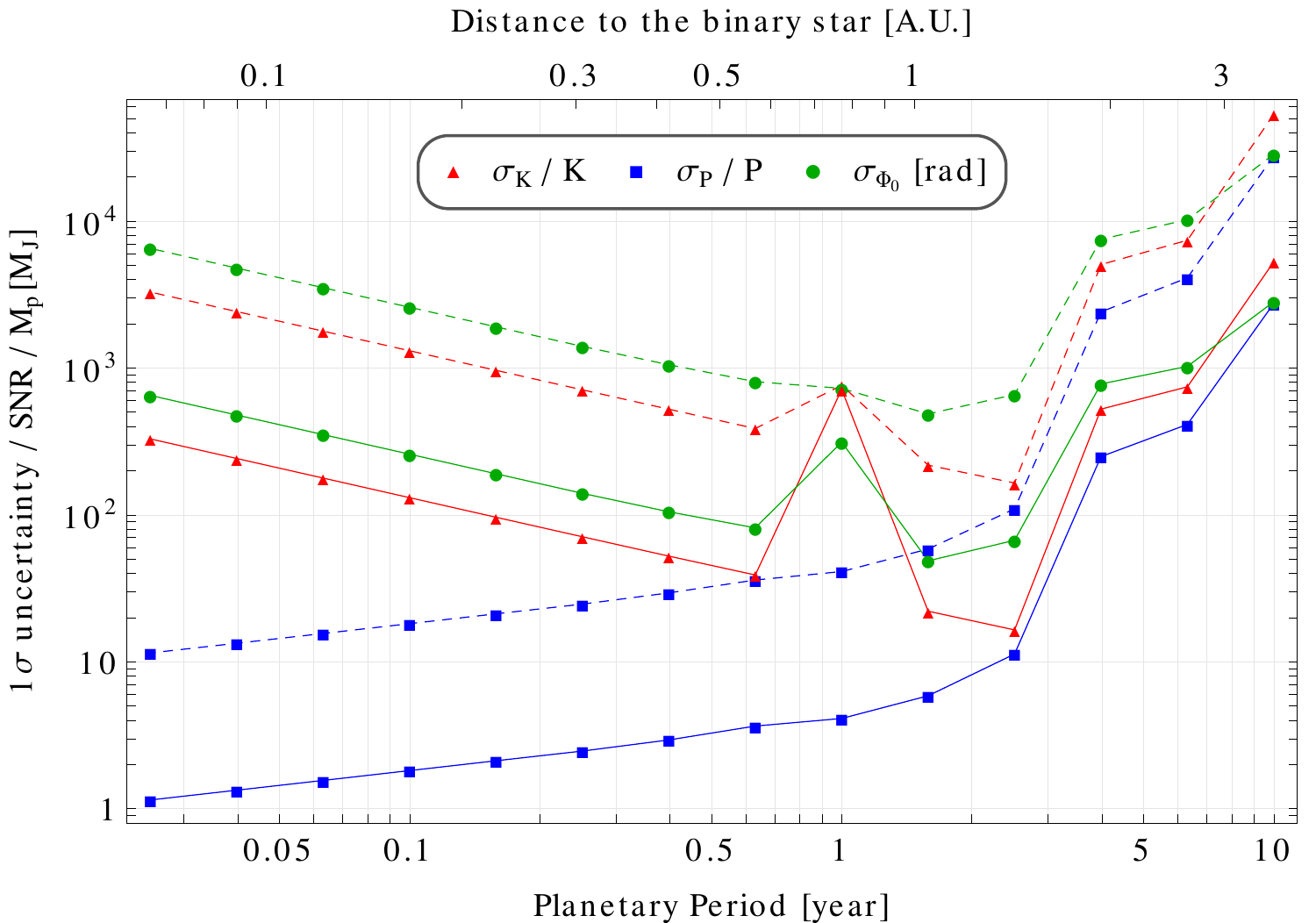}
\caption{
\textbf{LISA estimation of planetary parameters}.
Relative accuracy on $K$ and $P$, and absolute accuracy on $\varphi_0$, for a LISA measurements of planetary orbital parameters.
Solid curves denote DWD with $f_0 = 10$ mHz, while dashed curves are for $f_0 = 1$ mHz.
All results have been scaled with respect to the DWD SNR and the planetary mass measured in Jupiter masses.
The peak at one year is caused by degeneracies in the GW signal appearing when the planetary orbit is an exact multiple of LISA's orbital period around the Sun.
}
\label{fig:error_estimates}
\end{figure}

The three parameters of the circumbinary planet (CBP), which can be recovered with LISA, are the planetary period $P$, its initial phase $\varphi_0$ and the parameter
\begin{equation}
 K = \left(\frac{2\pi G}{P}\right)^{\frac{1}{3}} \frac{M_p}{(\Mb+M_p)^{\frac{2}{3}}} \sin i  \,,
 \label{eq:K}
\end{equation}
which depends on the CBP mass $\Mp$, its orbit inclination $i$, and the binary total mass $M_b$.
In Fig.~\ref{fig:error_estimates} we show how the 1$\sigma$ relative uncertainties on $K$, $P$, and the 1$\sigma$ absolute uncertainties on $\varphi_0$
(in radians),
vary as a function of $P$.
We consider two values for the GW frequency emitted by a DWD: a representative frequency at $f_0 =$ 1 mHz (dashed lines), at which the majority of DWDs detected by LISA are expected\cite{Korol2017}, and a higher frequency $f_0 =$ 10 mHz (solid lines), where events with higher signal to noise ratio (SNR) will be detected.
All numbers are linearly scaled with respect to the SNR of the DWD detected by LISA and the mass of the CBP.
This implies that in order to find the precision with which the parameters are measured, the numbers reported in Fig.~\ref{fig:error_estimates} must be divided by the SNR of each individual DWD event and by the CBP mass, measured in $\Mj$ (Jupiter's mass).

Fig.~\ref{fig:error_estimates} shows that the error estimations on $K$ and $\varphi_0$ are better for planetary periods $P$ comparable to the LISA nominal lifetime.
For shorter planetary periods the errors on $K$ and $\varphi_0$ smoothly decrease with the increasing of $P$, while for planetary periods longer than LISA's nominal duration the precision steeply worsen by few order of magnitude.
On the other hand, the uncertainty on $P$ increases smoothly with the planetary period till the LISA nominal mission lifetime, after which it rapidly worsen similarly to the behaviour of $K$ and $\varphi_0$.
An explanation for the different behaviour of $P$, as well as for the appearance of the peak at one year, is provided in the Methods section.

Analogously to EM radial velocity techniques, we cannot recover the mass of the companion directly from the three GW parameters $ K,\ P,\ \varphi_0 $, implying that we have no mean to know if the perturbing object is a planet, a star or a different object.
Nevertheless,
if both $K$ and $P$ are measured with sufficient precision, then additional EM information on the total DWD mass and on the companion's orbit inclination, can help determining its mass through Eq.~\eqref{eq:K}.
It is furthermore possible to obtain some estimates on the allowed $\Mp$ range if also the chirp mass of the binary is measured with LISA, which is usually the case at least for high frequency DWDs\cite{Takahashi:2002ky}.
In fact, by assuming that the symmetric mass ratio $\eta$ of DWDs detected by LISA cannot be lower than a certain value, we can derive upper and lower bounds for the total DWD mass $M_b$, and from this range of values we can estimate $\Mp \sin(i)$ through Eq.~\eqref{eq:K}.
Since from GWs alone we do not obtain any information on $i$,
we can only derive lower limits for $\Mp$ by considering $\sin(i) = 1$.
Values of $\sin(i)$ different from 1 will yield higher values of $\Mp$. 
If the lower mass limit estimated in this way is below 13 $\Mj$ (the deuterium burning limit), then the companion object is probably an exoplanet which would need to be confirmed by EM follow-ups (cf.~Sec.~\ref{sec:EM_analysis}).
On the other hand, if the lower bound of this estimate is above 13 $\Mj$, then we can confidently exclude the possibility that the third orbiting object is a planet.
Furthermore, if we also set a minimum allowed value for $\sin(i)$, we can find an upper bound on $\Mp$ which would confirm the presence of an exoplanet from the GW signal alone if it does not exceed 13 $\Mj$.
In this case no EM counterpart would be needed to confirm the GW detection of a CBP.

\begin{figure}[t]
\centering
\includegraphics[width=0.8\columnwidth]{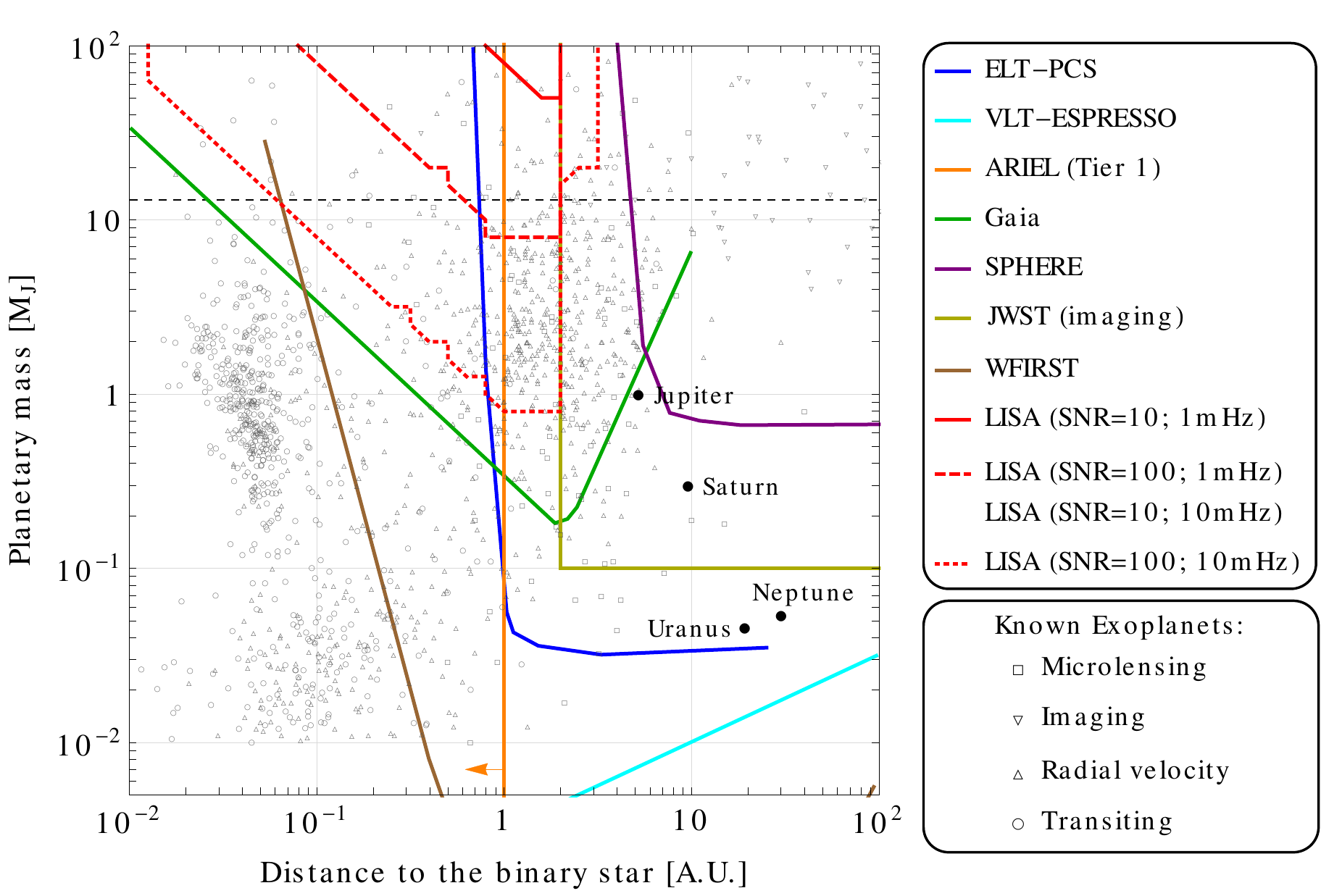}
\caption{
\textbf{Selection functions of both LISA and main EM exoplanetary projects}.
Selection functions of LISA (gravitational waves, \textit{red} lines): planetary mass $\Mp$ versus distance from the (binary) star.
The DWD GW frequency $f_0$ and the LISA SNR 
are marked in the legend. 
The typical selection curves of other exoplanet detection experiments are plotted for comparison: 
$Gaia$ (astrometry, \textit{green}) line\cite{Casertano2008,Sozzetti2014} is measured for a 0.5 $\Msun$ M dwarf at 25 pc; VLT-ESPRESSO (RV, \textit{dashed blue}) line\cite{Pepe2014} is measured for a velocity amplitude of 10 cm s$^{-1}$ for a 0.8 $\Msun$ star; ARIEL (transiting, \textit{orange}) line\cite{ARIEL} is based on the tier 1 sample; WFIRST (microlensing, \textit{brown}) line\cite{Penny2018} corresponds to the 3$\sigma$ detection; ELT-PCS (imaging, \textit{solid blue}) line\cite{Lagrange2014}, JWST (imaging, \textit{lime}) line\cite{Beichman2010}, SPHERE (imaging, \textit{purple}) line\cite{Lagrange2014}. Please note that some of the lines overlap.
Gray markers corresponds to the currently known exoplanets as reported in \textbf{\color{blue} exoplanet.eu}.
The dashed horizontal line marks the deuterium burning limit at $\Mp = 13 \Mj$.} 
\label{fig:MassVSsep}
\end{figure}

Given the discussion above, we are particularly interested in systems for which both $K$ and $P$ can be measured with a relative statistical uncertainty better than some accuracy value which we arbitrarily set to 30\%.
In what follows we will define a \textit{detection} of a CBP with GWs if our estimated precision on both $K$ and $P$ is better than this number. 
Fig.~\ref{fig:MassVSsep} shows the region in the mass-separation parameter space where LISA will have the possibility of detecting CBPs, according to the definition of detection above.
From the figure we see that LISA will be more effective at observing exoplanets with a separation from the DWD between roughly 1 and 2 AU.
Depending on the SNR and frequency of the DWD, LISA will be able to detect exoplanets down to $\sim$1 $\Mj$ and will be efficient in the range between 0.01 and 3 AU, roughly.

Although the occurrence rate of CBPs around DWDs is presently unknown, if these planets physically exist, and they orbit even a few percent of DWDs detectable by LISA, we will observe up to hundreds of new exoplanets, assuming they are sufficiently massive to be heard.

Our results are obtained for binaries composed by two WDs,
each of mass $\Ms = 0.23 \Msun$ (chirp mass $M_c = 0.2 \Msun$).
For DWDs with higher (chirp) masses the perturbation due to a planet will in general be weaker, although the SNR of the binary will be higher for equal distance, symmetric mass ratio $\eta$ and GW frequency.
The same reasoning applies to different types of stellar binaries visible with LISA, such as neutron star-neutron star and WD-neutron star binaries, for which sufficiently massive CBPs could be observed as well.

In our analysis we assume for simplicity no eccentricity for both stellar and planetary orbits.
Note, however, that the eccentricity of the planetary orbit constitutes another measurable parameter with LISA, if it differs significantly from zero\cite{Robson2018}.
For eccentric DWDs, which may form in the presence of a hierarchically bound third companion\cite{Thompson:2010dp,Seto:2013wwa}, it might also be possible to measure the individual masses of the stars directly from GW observations\cite{Valsecchi:2011mv}, yielding a more accurate and precise estimation of the CBP mass without the need of complementary EM observations.

We moreover assume the DWDs to be detached, namely with no accretion.
LISA is expected to detect also few thousands accreting DWDs\cite{Kremer:2017xrg}, for which however it might be difficult to directly measure the chirp mass with GWs and to disentangle an eventual CBP signal from accreting effects.
Nevertheless, individual masses can instead be recovered with joint GW-EM observations, for instance with LISA-$Gaia$ for which $\sim$50 such events can be characterised\cite{Breivik:2017jip}.
Moreover accreting DWDs are expected to emit specific EM signatures which might facilitates the identification of an EM counterpart\cite{Nelemans:2003ha}, enhancing the multi-messenger potential of these systems.

In our analysis we assume a 4 years nominal LISA mission duration\cite{LISAcallpaper}.
However, for a maximal extension of 10 years, we expect the number of DWD detections to roughly double\cite{Korol2017}, and the SNR of each individual DWD to increase as the square root of time\cite{Cutler1998,Takahashi:2002ky}.
This translates into both a higher number of CBPs detections with GWs, and a more precise characterisation of the systems already observed in 4 years.
A 10 years LISA mission will also be more sensitive to CBPs with periods between 4 and 10 years.
In fact we expect the rapid deterioration present at periods longer than 4 years of the accuracy with which the planetary parameters are recovered (cf.~Fig.~\ref{fig:error_estimates}), to shift 
toward longer periods, in particular around 10 years.

Our results are always rescaled with the SNR of the DWD, implying that all effects due to the distance of the GW source from the Earth have been factored out.
LISA will detect DWDs at all distances within the Milky Way, and even at intergalactic distances up to M31\cite{Korol2018} (Andromeda galaxy).
Although only systems in the solar neighbourhood ($d_* \lesssim$ 3 kpc) can be repeatedly observed with joint GW and EM observations, farther CBPs can be detected by GW alone.
Given that no EM detection of extragalactic planets bound to their star(s) has been confirmed yet (see \cite{Sumi2011,Dai2018} for detection of population of unbound planets), LISA will open a new window for the quest of extra-galactic exoplanets through GW astronomy.\\
The other advantage of having produced SNR rescaled results, is that we can directly make predictions not only for different DWD systems detected with the same LISA configurations, but also assess the potential for exoplanetary searches of other LISA-like missions.
For instance, for DWDs with SNR=100 and $f_0 = 10$ mHz LISA will be able to hear CBPs of $\Mp \geq 1\, \Mj$. 
Some of these higher frequency DWDs will be measured with SNR $\sim$ 1000, meaning that, for these specific systems, LISA will be able to detect CBPs down to 30 M$_\oplus$ ($\sim 0.1\, \Mj$).
Similarly, a future more sensitive GW observatory, which can improve LISA SNR values of at least one order of magnitude, 
will be able to observe CBPs down to few Earth masses. 
These CBPs might as well be located in the habitable zone (HZ) of the DWDs ($a\sim$ 0.014 AU, to account for the double luminosity)\cite{Agol2011}.


\section{Synergy with EM observations and implications}
\label{sec:EM_analysis}

Observing the EM counterpart is important to fully characterise the planetary system. 
When a planet is first heard by GWs (cf Sec. \ref{sec:results}), we can precisely determine $\Mp$
only if the total DWD mass is measured with EM observations.
Moreover when the mass of each component in the binary is determined, then it is possible to estimate their radius using mass-radius relations\cite{Veras2016}.
Recent estimates\cite{Korol2017} suggest that out of the $25\times 10^3$ DWDs expected to be detected in 4 years by LISA, only up to a maximum of 100 will be observed with instruments such as Gaia and LSST, although upcoming large ground-based facilities might well improve these numbers.
We verified also that once the EM counterpart has been spotted, the LISA accuracy on $K$ and $P$ parameters improves only marginally by fixing the sky location. 

On the other hand, the CBP itself has its typical signature that can be spotted in the EM
by various well-known detection techniques (cf. Fig. \ref{fig:MassVSsep}), which on average mainly observe systems relatively close to us.
Transit, radial velocity (RV), direct imaging, and astrometry methods cover distances up to $d_{*}\sim$ 3 kpc, while gravitational
microlensing goes up to $d_{*}\sim$ 8 kpc.
Among these, while the transit and RV methods are affected by stellar activity, direct imaging is limited by the angular resolution of the instrument.
Microlensing events are not reproducible and astrometry is severely limited by the timespan of the survey.
The CBP detection method by GWs proposed here is conceptually similar to RV, but has the advantages that it can observe everywhere in the galaxy, it is not affected by the activity of the stars, and it is does not need any observational pointing.\\
If we assume that the planetary system complies the requirements specific for the EM detection (e.g. magnitude and system configuration), the synergy with GWs
allows us to resolve the circumbinary system in term of periods, masses (of the stars + planet), and inclination of the orbits.
We refer to the Methods for more details on the parameters that can be retrieved in the specific synergy cases.

We note that the orbit of a CBP is stable when the semi-major axis $a$ is approximately $a\ \gtrsim 4.5$ the binary period $P_b$\cite{HolmanWiegert1999, PilatLohinger2003},
which in our specific case corresponds to a minimal separation of $a_{min} \sim $ 0.0009 AU and $a_{min} \sim$ 0.004 AU for the binary period $P_b$ = 30 min and $P_b$ = 1 hour, respectively. 
These distances are well below the minimal separation considered for CBPs in this work, since we always assume $P \gg P_b$ in order for higher order orbital effects, e.g.~Kozai-Lidov resonances, to be negligible.

WDs are common as Sun-like stars and may also provide a source of energy for planets for gigayear (Gyr) durations.
Cool white dwarfs ($T_{\rm eff}$ \textless  6000 K), have an HZ that endures for up to 8 Gyr\cite{Agol2011} located at only few solar radii from the star.
This HZ regularly moves inwards with the star aging, and hence cooling\cite{Agol2011}. 
Planets in the HZ of a WD must have consequently migrated inwards after the stellar giant phase\cite{DebesSigurdsson2002,Livio2005,Faedi2011}.
The detection of such objects, or any close-in object not necessarily in the HZ, would help constraining migration theories of planets around post-common envelope binaries. 
For a binary case and with hotter DWDs (both at $T_{\rm eff}$ \textgreater 6000 K), the HZ scales outwards to account for the energy contribution of both stars and their temperatures.

It has been shown that, after one inner binary common envelope (CE) phase, CBPs can either be ejected by the system\cite{VerasTout2012, Veras2016}, or survive the CE phase(s) while undergoing an orbital expansion/shrinkage, generally coupled with an increased eccentricity\cite{Kostov2016}. The kind of binary evolution, i.e., the tidal evolution, the energy transfer between the CE and the inner cores, and the mass-loss rate, is directly responsible for the fate of their planetary companion.
Consequently, the detection and characterisation of CBPs around DWDs is crucial for enabling a 
comparison with known existing circumbinary systems, and for pinning down the planetary evolution phases, valid all over the Galaxy, while their binary evolves. 
An example is the circumbinary system Kepler 1647\cite{KostovKepler2016}(gas giant orbiting two solar-mass stars of F and G type), which could, under a specific theoretical scenario, become a WD-WD pair that keeps a bound exoplanet after two CE phases\cite{Kostov2016}.

As previously mentioned,
a 10 years mission lifetime can detect and characterise exoplanets on wider orbits, making LISA even more compatible with 
EM direct imaging techniques, and specifically with large ground-based cameras (e.g., ELT-PCS, cf.~Fig.~\ref{fig:MassVSsep}) or the successor of JWST.
Imaging of CBPs around DWDs can be used to test the presence of
a second-generation exoplanets in the outer regions of planetary system, and consequently to provide constraints on migration theories. 
Emission spectra of these objects will furthermore allow us to estimate its temperature and the main molecular component of its atmosphere, making direct connections to chemical element distributions in the WDs atmosphere.
This would also permit to better understand the observed
WD pollution effect\cite{Veras2016}. On another hand, if an existing CBP accretes mass after a CE stage, it becomes brighter, further decreasing the already low planet-to-WDs contrast, meaning that also first-generation, more mature exoplanets, can be imaged.

Another peculiarity of these stars is their size. Their small radius, $R_{WD} \approx$ 1 $R_\oplus$, makes them excellent targets for detecting transiting exoplanets, and in particular Earth-like planets\cite{Agol2011}. 
However, Jupiter-like planets, which are a factor  $\sim 7$ larger than a typical WD, generates a complete eclipse during their transit, fully occulting the stellar light. 
This facilitates their detection, but prevents atmospheric studies on the exoplanet in question through transmission spectroscopy. For specific configurations, though, atmospheric feature information could be retrieved by using an external background source.  
If the planet is in close-in orbit and have an exomoon (assuming it is not removed through an evection resonance while undergoing the planetary migration phase\cite{Spalding2016}), then transit spectroscopy will allow to study the atmosphere of the satellite, due to the small 
size of WDs.
For instance, an 
Europa transit ($R_E \sim$ 0.24 R$_\oplus$) in front of a typical WD ($\Ms = 0.6\ \Msun$, $R_{WD} = 1.57$ R$_\oplus$) yields a decrease in brightness of approximately $\sim 5 \%$.

\section{Conclusion}

We have presented an original observational method which employs gravitational waves to detect exoplanets.
The conceptual idea is similar to the Radial Velocity detection technique, but it is notably not affected by stellar activity.
Our results show that LISA will allow us to verify the existence of exoplanets orbiting detached white dwarfs binaries in our Galaxy, as well as in other nearby galaxies.
The discovery of  such objects will statistically increase the actual sample of post-main sequence planets, filling an area of the planetary HR diagram currently not explored. 
Such a population will be unbiased and valid all over the Milky Way. 
Specifically,  LISA will provide observational constraints on both planets that can survive two common envelope stellar evolution phases, and on a possible second-generation (or maybe third?) planet population.
Depending on the SNR of the observation, LISA will have the potential to detect super-Earth exoplanets with mass down to 30 M$_\oplus$. 
On the other hand, in a scenario where LISA will not detect any of these circumbinary planets anywhere in the Milky Way, we will still be able to set strong unbiased constraints on planetary evolution and dynamical theories.

Given that in the next few years the Transiting Exoplanet Survey Satellite (TESS) is expected to observe $\sim$ 500,000 bright eclipsing binaries\cite{Quarles2018}, a hefty increase in the number of observed P-type circumbinary planets, mostly around main sequence binaries, is expected.
These discoveries are important for collecting a more robust statistics about such systems in the solar neighbourhood, offering the unique opportunity to improve our understanding of the planets' dynamics in binary systems. 
This is particularly relevant for close ($\lesssim$10 AU) binary systems, the ancestors of the systems observable by LISA, whose stellar evolution is qualitatively and quantitatively different from wide orbit binaries.
Together with these upcoming evidences, a further development in the planet-star interaction and binary evolution theoretical studies (e.g.~\cite{Kostov2016, Mustill2013,Portegies2013}) needs to be performed 
in the forthcoming years, in order to thoroughly assess the potential of LISA to detect exoplanets, especially before 
the definitive ESA approval of the mission design in the early 2020s\cite{LISAcallpaper}.

The characterisation of LISA circumbinary exoplanets, by both GW and EM observations, will deepen our knowledge on planet-star interaction during the critical phases of stellar evolution, and will enable us to find the connecting thread that runs from formation to the very end of an extrasolar planet.
GWs are probably not the answer to the ultimate question of life, the universe, and everything, but they might after all constitute the key to find \textit{Magrathea}\cite{Magrathea}.

\section*{Acknowledgements}

It is a pleasure to thank Emanuele Berti, Alessandra Buonanno, Valeryia Korol, Pierre-Olivier Lagage, Antoine Petiteau, Elena Maria Rossi and Giovanna Tinetti for their suggestions and comments.
C.D.~acknowledges support from the LabEx P2IO, the French ANR contract 05-BLAN-NT09-573739.

\newpage

\section*{Methods}
\label{sec:methods}

\subsection*{Dynamical model and modified GW phase}

The three-body systems we are interested in are composed by a DWD emitting GWs in the LISA frequency band and a planet orbiting the DWD on an outer orbit.
We assume that the separation between the planet and the DWD is much greater than the separation between the two stars forming the DWD.
Moreover the period of the planetary orbit is always assumed to be much longer than the period of the inner DWD orbit.
These assumptions imply that in first approximation these three-body systems can be treated as two separate two-body problems.
In particular the internal orbit of the DWD is not perturbed by the CBP, and the planet and the CoM of the DWD are orbiting each other on Keplerian orbits.
For the sake of simplicity in our investigation we consider both these orbits as circular.
The extension to elliptic orbits should be straightforward and will be left for future more in depth investigations.

\begin{figure}[p]
\centering
\includegraphics[height=0.3\textwidth]{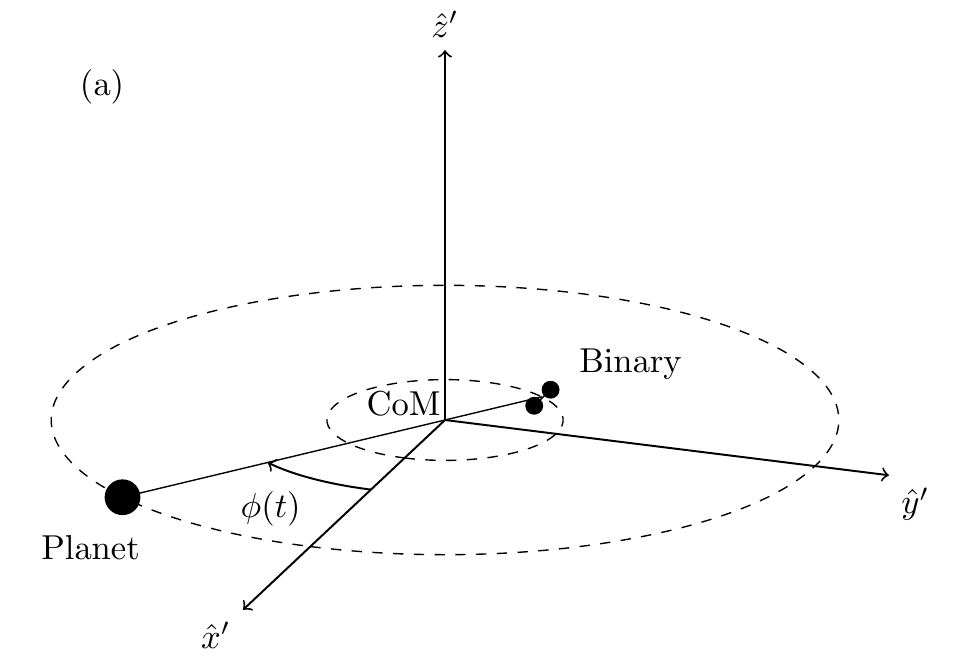}
\hspace{1cm}
\includegraphics[height=0.3\textwidth]{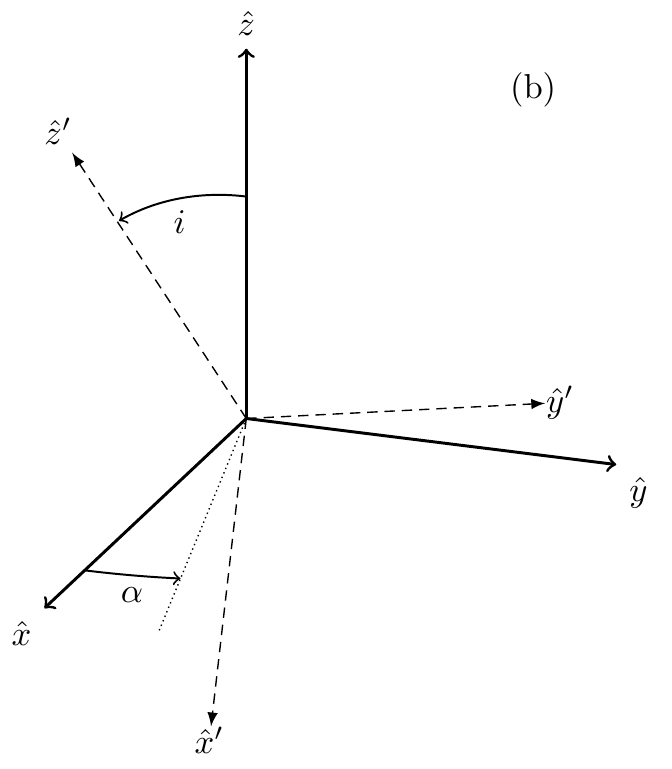}
\caption{
\textbf{Geometry of the DWD-CBP three-body system}.
Left panel: source reference frame (primed coordinates).
Right panel: observer reference frame (non-primed coordinates) with the source reference frame rotated.
}
\label{fig:geometry}
\end{figure}

Given these assumptions, in the $(x',y',z')$ reference frame where the direction $\hat{z}'$ is perpendicular to the planetary orbital plane (cf.~Fig.~\ref{fig:geometry}), the distance vector between the planet and the CoM of the DWD is given by
\begin{equation}
 \mathbf{r}(t) = ( R \cos\varphi(t), R \sin\varphi(t), 0 ) \,,
\end{equation}
where $R$ and $\varphi$ are
\begin{equation}
 R^3 = G (M_b + M_p) \left( \frac{P}{2 \pi} \right)^2 \quad\mbox{and}\quad \varphi = \frac{2 \pi t}{P} + \varphi_0 \,,
\end{equation}
with $M_b$, $M_p$, $P$ and $\varphi_0$ the total mass of the binary, the mass of the planet, the period and the initial phase of the planetary orbit, respectively.
To find the motion in the $(x,y,z)$ reference frame of a general observer whose $\hat{z}$-direction points toward the source, we apply two rotations: a first rotation by $i$ (the inclination angle between $\hat{z}'$ and $\hat{z}$ - the line of sight) to bring the $\hat{z}'$ direction to point toward the $\hat{z}$ direction, and a subsequent second rotation by $\alpha$ around the $z'$ axis (which after the first rotation coincides with the $z$ axis) making $\hat{x}'$ and $\hat{y}'$ to point in the same directions of $\hat{x}$ and $\hat{y}$ (see Fig.~\ref{fig:geometry}).
Note however that the second rotation around $\hat{z}$ is degenerate with a phase redefinition of the planetary orbit, i.e.~it is degenerate with $\varphi_0$ and can thus be ignored.
This holds as long as one assumes the CBP orbit to be circular and ceases to be true as soon as non zero eccentricity is considered.
The motion in the reference frame of a general observer is thus given by
\begin{equation}
 \mathcal{R}(i)\mathbf{r} = ( R \cos\varphi(t), R \cos i  \sin\varphi(t), -R \sin i  \sin\varphi(t) ) \,,
\end{equation}
where $\mathcal{R}$ is a rotation matrix by the angle $i$.

For what concerns our scopes, we are interested in the circular motion of the CoM of the DWD around the common CoM of the three-body system.
The distance vector connecting the DWD CoM to the three-body system CoM is
\begin{equation}
 \mathbf{r}_b = \frac{\Mp}{\Mb+\Mp} \mathcal{R}( i )\mathbf{r} \,.
\end{equation}
We are only interested in the motion along the line of sight from the observer to the system\footnote{For any practical purposes the direction pointing from the observer to the CoM of the three-body system and the direction pointing from the observer to the DWD CoM will be assumed to coincide.}.
Since the z-axis of the observer is aligned along the line of sight direction, the z-component of the motion, given by
\begin{equation}
 z_b = - \frac{\Mp}{\Mb+\Mp} R \sin i  \sin\varphi(t) \,,
\end{equation}
is the one we need to consider.
The velocity of the DWD CoM along the line of sight is then given by
\begin{equation}
 v_{z,b} = - K \cos\varphi(t) \,,
\end{equation}
where we defined the parameter
\begin{equation}
 K = \left(\frac{2\pi G}{P}\right)^{\frac{1}{3}} \frac{M_p}{(\Mb+M_p)^{\frac{2}{3}}} \sin i  \,.
 \label{eq:gamma}
\end{equation}

In the reference frame comoving with its center of mass, the DWD emits almost monochromatic GWs at some specific frequency $f_{GW}$.
Since this frequency is changing on time scales much longer if compared to the observational time scale, an expansion around the initial observed frequency will suffice in describing their time evolution\cite{Takahashi:2002ky}.
We can thus assume
\begin{equation}
 f_{GW}(t) = f_0 + f_1 t + \mathcal{O}(t^2) \,,
\end{equation}
where $f_0$ is the initial observed frequency, $f_1$ is the time derivative of $f_{GW}$ evaluated at the initial time and we neglected second and higher order terms.
The GW frequency in the observer reference frame changes instead due to the Doppler effect which, as long as the dynamics is non-relativistic, gives
\begin{equation}
 f_{obs}(t) = \left( 1 + \frac{v_{z,b}(t)}{c} \right) f_{GW}(t) \,.
\end{equation}
Finally the phase at the observer of the GW can be obtained integrating the frequency $f_{obs}$:
\begin{equation}
 \Psi_{obs}(t) = 2\pi \int f_{obs}(t') dt' + \Psi_0 \,,
\end{equation}
where $\Psi_0$ is a constant initial phase.
This is thus the GW phase measured by LISA from which the parameters characterizing the DWD and the planetary perturbation can be extracted.

\subsection*{LISA parameter estimation}

The three arms of LISA constitute a pair of two-arm detectors outputting two linearly independent signals $h_{I,II}(t)$.
Assuming that the noise in each independent channel is stationary and Gaussian, these two signals can be expressed in the common amplitude-and-phase form as\cite{Cutler1998}
\begin{equation}
 h_{I,II}(t) = \frac{\sqrt{3}}{2} A_{I,II}(t) \cos\left[ \Psi_{obs}(t) + \Phi^{(p)}_{I,II}(t) + \Phi_D(t) \right] \,,
 \label{eq:obs_wf}
\end{equation}
where 
\begin{align}
 A_{I,II}(t) &= \left[ A_+^2 F_{I,II}^{+2}(t) + A_\times^2 F_{I,II}^{\times 2} \right]^{1/2} \,,\\
 \Phi_{I,II}(t) &= \tan^{-1}\left(-\frac{A_\times F_{I,II}^\times(t)}{A_+ F_{I,II}^+(t)}\right) \,, \\
 \Phi_D(t) &= \frac{2\pi f_{obs}(t) R_{\rm Earth}}{c} \sin\theta_S \cos\left(\bar\phi_0 + \frac{2\pi t}{P_{\rm Earth}} - \phi_S\right) \,. \label{eq:LISA_orbit_3}
\end{align}
In these expressions $R_{\rm Earth} = 1$ AU and $P_{\rm Earth} = 1$ year are the mean distance from the Sun and the orbital period of the Earth.
$F_{I,II}^{+,\times}$ are the antenna pattern functions depending on the source angular position $(\theta_S,\phi_S)$, the orientation of its orbit $(\theta_L, \phi_L)$ and the LISA configuration.
The quantities $A_{+,\times}$ are instead constant amplitudes which depend on the physical parameters and orientation of the source.
Eqs.~\eqref{eq:obs_wf}--\eqref{eq:LISA_orbit_3} provide the GW signal measured by LISA irrespectively of the sky location of the source, i.e.~for arbitrary values of $(\theta_S,\phi_S)$.
The full expressions for all the quantities appearing in these equations can be found in the literature\cite{Cutler1998,Takahashi:2002ky,Cornish:2003vj}.
We basically follow the set up described in\cite{Cutler1998}, adding to the GW signal the perturbation due to the CBP.

In order to extract information on the parameters characterizing the three-body system under consideration, we employ matched filtering techniques\cite{Cutler1998,Takahashi:2002ky}.
We assume that the GW signal measured by LISA has the time dependent expression given in Eq.~\eqref{eq:obs_wf}.
This is characterized by 11 parameters, namely $\ln(A), \Psi_0, f_0, f_1, \theta_S, \phi_S, \theta_L, \phi_L, K, P, \varphi_0 $ which we collectively denote by $\lambda_i$.
The SNR of the signal can be computed as
\begin{equation}
 {\rm SNR}^2 = \frac{2}{S_n(f_0)} \sum_{\alpha = I,II} \int_0^{T_{obs}} dt\, h_\alpha(t) h_\alpha(t) \,,
 \label{eq:SNR}
\end{equation}
where $T_{obs}$ is LISA observational time and $S_n(f_0)$ is LISA one-sided spectral density noise computed at $f_0$.
In all computations we fix $T_{obs} = 4$ years in agreement with the nominal mission requirements\cite{LISAcallpaper}.
For all the systems we consider here, the SNR as defined in Eq.~\eqref{eq:SNR} is approximately the same as the SNR of an equivalent DWD without the perturbation of the planet.
For this reason in our analysis we freely compare this SNR with values reported in the literature for non perturbed DWDs.
For GW signals with high SNR, the uncertainties and correlations on the parameters can instead be estimated from the covariance matrix $\Sigma_{ij}$, given by the inverse of the Fisher information matrix
\begin{equation} 
 \Sigma_{ij} = \left< \Delta\lambda_i \Delta\lambda_j \right> = (\Gamma^{-1})_{ij} \,.
\end{equation}
The standard estimator for the statistical error on the parameter $\lambda_i$ is thus given by $\sqrt{\Sigma_{ii}}$.
The Fisher information matrix itself can be computed as
\begin{equation}
 \Gamma_{ij} = \frac{2}{S_n(f_0)} \sum_{\alpha = I,II} \int_0^{T_{obs}} dt \frac{\partial h_\alpha(t)}{\partial\lambda_i}\frac{\partial h_\alpha(t)}{\partial\lambda_i} \,.
 \label{eq:FM}
\end{equation}
Note that we are simplifying Eqs.~\eqref{eq:SNR} and \eqref{eq:FM} by considering $S_n(f_0)$ to be a constant and thus by taking it out of the time integral.
This approximation is justified for almost monochromatic signals whose frequency does not depart considerably from $f_0$.
Consequently in analogy with the works of\cite{Cutler1998,Takahashi:2002ky}, in our analysis we always replace $S_n(f_0)$ with the SNR of the source by using Eq.~\eqref{eq:SNR}.
This allows us to scale all results with the SNR of each event, without referring to any particular configuration of LISA.

The accuracy with which the parameters associated with the DWD orbit, namely $\ln(A)$, $\Psi_0$, $f_0$, $f_1$, $\theta_S$, $\phi_S$, $\theta_L$, $\phi_L$, can be observed by LISA, has already been investigated in several works\cite{Cutler1998,Takahashi:2002ky,Cornish:2003vj}.
For this reason we focus our analysis on exploring the possibility of measuring the additional parameters coming from the perturbation due to the planet, namely $ K, P, \varphi_0 $.
We moreover restrict the sampled parameter space by considering only specific values for some of the DWD parameters.
In analogy with the examples in\cite{Cutler1998,Takahashi:2002ky}, we set the orbital geometry of the DWD and its sky location by assuming
\begin{equation}
  \Psi_0 = 0; \quad \theta_S = \arccos(0.3); \quad \phi_S = 5; \quad \theta_L = \arccos(-0.2); \quad \phi_L = 4 \,.
\end{equation}
Different choices of the parameters above do not alter the qualitative conclusions derived in our investigation, although clearly the DWD SNR will change if these parameters change.
By considering representative values from the expected population of DWDs observed by LISA\cite{Korol2017}, we assume equal mass DWDs with a chirp mass of $M_c = 0.2\, \Msun$.
The value of $M_c$, together with the frequency $f_0$, yields $f_1$ as
\begin{equation}
  f_1 = \frac{96}{5} \pi^{8/3} f_0^{11/3} \left(\frac{G M_c}{c^3}\right)^{5/3} \,.
\end{equation}
The value of the amplitude $A$ is irrelevant for our analysis since we always report results in terms of SNRs.
The most favorable orientation for the detectability of the planetary perturbation on the GW signal consists in having $i = \pi/2$.
We set $i$ to this value, keeping in mind that different inclinations will always be expected to give worse results.
Analogously we set $\varphi_0 = \pi /2 $.
The results might change if other values of $\varphi_0$ are chosen, but at least for periods shorter than the LISA mission duration this choice should not affect them.
The remaining parameters, namely $f_0$, $P$ and $K$, are varied in order to explore the parameter space.
Since we analyse only systems for which the planetary inclination $i$ and DWD total mass $M_b$ (through $M_c = 0.2 \Msun$ and $\eta = (M_c / M_b)^{5/3} = 1/4$) are fixed, the choice of the planetary mass directly determines the value of $K$.
For this reason all results are reported by referring to the planetary mass $M_p$.
We provide results for a LISA nominal mission duration of 4 years\cite{LISAcallpaper} and neglect the presence of a possible second planet or tertiary star, which are commonly found around binaries orbiting DWDs\cite{Tokovinin2006}.

\subsection*{Details of Fig.~\ref{fig:error_estimates}}

The precision on the CBP orbital parameters, as estimated with the methodologies exposed above, is reported in Fig.~\ref{fig:error_estimates}.

First of all we have verified that as long as our assumptions hold, errors on the planetary parameters change linearly as $\Mp$ changes.
The CBP mass can thus be factored out from the final estimation of the measurement precision on the parameter, similarly to what we have done with the SNR.
This is why in Fig.~\ref{fig:error_estimates} we present results scaled by both the DWD SNR and the CBP mass, and why in Fig.~\ref{fig:MassVSsep} the LISA selection function for SNR = 100 and $f_0 = 1$ mHz coincides with the one for SNR = 10 and $f_0 = 10$ mHz.

The behaviour of the uncertainty on $P$ in Fig.~\ref{fig:error_estimates} differs from the behaviour of $K$ and $\varphi_0$.
An explanation for this can be found by looking at the form of the signal in the Fourier domain (cf.~Fig.~5 of\cite{Robson2018}).
In fact the spread of the signal over different frequency bins due to the Doppler modulation induced by the CBP, directly determines the orbital period of the planet.
This spread is easier to measure with respect to other details of the GW signal which are used to determine $K$ and $\phi_0$, together with the other DWD parameters.
This implies that the CBP period $P$ will be always well measured as long as it is shorter than the LISA observational time.
For longer CBP periods this spread is less effective since only a fraction of the orbit is observed and the signal has no time to be fully spread, which in turn implies that $P$ is recovered with less precision.
Of course the longer the CPB period, the lower is the spread and thus the less precise the measurement of $P$.
Note that on top of the frequency spread given by the Doppler modulation due to the CBP, there is also the usual spread given by the orbital motion of LISA around the Sun.
It is however easy to distinguish the two spreads since we know the orbital period of LISA is exactly one year.
Similarly, the peak at one year on $K$ and $\varphi_0$ is due to the degeneracy between the motion of the DWD around the three-body CoM and the motion of LISA around the Sun.
We have indeed checked that other smaller peaks appear at multiples of one year, confirming that this effect is due to the degeneracy between the two orbital motions.

\subsection*{Synergy with radial velocity}

The radial velocity (RV) technique allows for the determination of the planetary orbital period, $P$, the semi-amplitude of the radial velocity curve, $K$, and the eccentricity, $e$ (which we did not take into account in this work), the longitude of the periastron, $\omega$, and the time of periastron passage $T_0$.
The drawback of such technique is that the RV signal can be strongly biased by the presence of stellar activity, causing a false positive detection of a planetary companion. 

Note that RV also allows for the determination of the stellar mass.
In the case of spectroscopic binaries (and if both stars are bright enough), each resolved stellar spectrum enables the estimation of the effective temperature and surface gravity, and consequently the determination of the stellar mass by using the WD mass-radius relations\cite{Althaus2005,Renedo2010}.
In the case of eclipsing spectroscopic binaries it is possible to solve for the mass of each member of the binary straight away.

The VLT/Echelle SPectrograph for Rocky Exoplanets and Stable Spectroscopic Observations (ESPRESSO) is designed to explore a new mass domain, corresponding to rocky planets down to the Earth mass in the habitable zone of solar-type stars, with a RV precision down to 10 cm s$^{-1}$ level\cite{Pepe2014}. 
Its selection curve is shown in Fig.~\ref{fig:MassVSsep}.
Such an instrument will be powerful enough to be able to follow up all possible potential exoplanets discovered by LISA, at least within its horizon (V magnitudes as faint as 20 to 21 in dark-sky conditions)\footnote{\url{https://www.eso.org/sci/facilities/paranal/instruments/espresso/ESPRESSO_User_Manual_P102.pdf}}.

For exoplanet detection RVs and GWs are conceptually similar methods, consequently the planetary parameters retrieved are the same.
However GWs are not affected by stellar activity and can  observe at larger distances.
RV is thus not expected to improve the CBP characterization already provided by GWs, but it will be extremely useful to measure the stellar masses and detect the EM counterpart, which would consequently confirm potential GW detections.

\subsection*{Synergy with transit}

The transit method\cite{TransitReview} alone allows for the determination of the planetary radius, $\Rp$, the orbital inclination, $i$, the semi-major axis, $a$.
Transit surveys, though, are affected by the issue of astrophysical false positives, which requires complementary observations to constrain the presence of a planet around a single star system. Conversely, a planetary detection in a circumbinary system is unambiguous due to 
the geometry (Fig. \ref{fig:geometry}) and dynamics of the system itself.
When the binary and the planetary orbits are coplanar\cite{PierensNelson2018,FoucartLai2013} the transits are only possible on eclipsing binaries, and when there is a misalignment, transits are still possible, but with gaps in the transit sequence and asymmetries in the transit profile\cite{MartinTriaud2014, Martin2017}.
Furthermore, because the planet is transiting a moving target, both
enhanced transit timing\cite{HolmanMurray2005,Armstrong2013} and transit duration variations\cite{Kostov2014,Liu2014} are expected\cite{Martin2017}.

Transit measurements, like RVs, are contaminated by stellar activity. However, a study on \textit{Kepler} data showed that WDs are photometrically stable to better than 1$\%$ on 1-hr to 10-d timescales\cite{Hermes2017}, key aspect for this kind of observations. 

We plot the selection function of the ESA-Atmospheric Remote-Sensing Infrared Exoplanet Large-survey (ARIEL; launch in 2028) mission\cite{ARIEL}, based on its Tier-1 sample. ARIEL goal is to investigate the atmospheres of several hundreds planets through transit and eclipse spectroscopy.

For the transiting hot-Jupiter whose mass and semi-major axis fall inside the LISA selection function (Fig. \ref{fig:MassVSsep}) 
it will be possible to determine $\Mp$, $i$ and $P$ (together with the eccentricity not considered here).

\subsection*{Synergy with astrometry}

Notably astrometry allows to determine all the orbital elements of a detected planet, 
 and yields a direct measurement of $\Mp/\Mb$ from which it is possible to determine the planetary mass if the binary mass is known.
Astrometry is sensitive to faint targets, long-period planets, and it is not affected by stellar activity. 
On the other hand this technique is limited by the timespan of the survey (which put limits on the period of exoplanets that can be detected) and, more importantly, by the distance of the system, which directly affect the precision on the individual measurements and hence the ability to resolve the system itself. Generally, at a fixed distance and stellar mass, the farther and the more massive the planet, the higher the astrometric signature.

No planets have been detected yet through this method, though $Gaia$, at the end of the nominal mission, will deliver a catalogue of 21000 ($\pm$6000) high-mass (1 - 15 $\Mj$) long-period planets at up to distances of $~$500 pc\cite{Perryman2014} and 500 circumbinary gas giants within 200 pc\cite{Sahlmann2015}.
$Gaia$ is also expected to detect tens or hundreds of planets around single WD ($\Mp$ \textgreater $\sim$ 1 $\Mj$) in long period orbits.
We report in Fig. \ref{fig:MassVSsep} the typical $Gaia$ selection function\cite{Casertano2008} updated 
with a most recent single-transit astrometric uncertainty model\cite{Sozzetti2014}.\\
In the case of joint GW-astrometric observations, we should be able to determine $\Mp$ from the combined measurements of $P$, $i$ and the ratio $\Mp/\Mb$.

\subsection*{Synergy with direct imaging}

Direct imaging of extrasolar planets is currently extremely challenging, especially at close projected separations, due to the high planet-to-star contrast.
Consequently, the method is currently sensitive to young giants self-luminous planets, orbiting at wide separations around low-luminous stars. \\
With only one epoch it is possible to measure the planetary angular separation, with more epochs over much longer timescales we can directly retrieve, $i$, $a$ and $P$\cite{Lagrange2014}. 
Other parameters such as $\Mp$, the actual effective temperature and gravity, are derived from the observed photometry using age-dependent relationships, meaning that an error on the age of the system translates in a large error on $\Mp$ and $\Rp$.\\
The advantage of imaging WDs is that such objects are natural clocks and their age, well constrained, can be derived straightforward from their luminosity\cite{Winget2008}. Obviously this would imply to be able to resolve both components in the binary system.
Furthermore, WDs are $\sim 10^4$ fainter than their progenitors, and the contrast ratio between DWDs-CBP
is lower after each CE phase.

No imaging of any post-common envelope binaries has been done yet, though gas giants have been found around a couple of these kinds of
systems with eclipse-time variations\cite{Qian2012, Beurmann2013}. 
These planets have large separations and can be imaged with upcoming large ground-based telescopes such as the ELT.  
The ELT will also allow us to perform spectroscopic observations of an unresolved planet and its exomoon(s)- (whether existing), 
and resolve the emitted combined light when the moon is hidden by the planet (i.e., during the moon eclipse).

We plot in Fig. \ref{fig:MassVSsep} the selection curve for ELT-PCS\cite{Lagrange2014} and SPHERE\cite{Lagrange2014} for comparison. 
While with SPHERE the synergy with LISA is inexistent (at the planetary mass level), the synergy with PCS covers mostly planets within $a$ = 1-2 AU,
whose $\Mp$ varies in the 1-13 (and above) $\Mj$ range.
If both the DWD total mass and the inclination $i$ are measured through direct imaging, then a joint observation with LISA will yield the planetary mass $M_p$.

One of the advantage of this technique is that it is not affected by the variability of the central star, it is though limited by the distance of the CBP system and by the angular resolution of the instruments.

\subsection*{Synergy with gravitational microlensing}

Gravitational microlensing alone provides star-to-planet mass ($\Ms/\Mp$) ratio and projected planetary separation in units of Einstein ring. 
However, it allows to retrieve also the distance $d_{*}$ of the system, $\Ms, \Mp$ and the semi-major axis $a$ in physical units, when working in synergy with adaptive optics observations and/or variations of the alignment between the foreground-background stars, due to a parallax effect. 
Microlensing detections are doable even for small planetary masses, and are more suitable for targets located between the centre of the galaxy and the observer. 
The downside is that every event is rare and unique, hence not reproducible.

Currently, only one circumbinary system has been discovered using gravitational microlensing\cite{Bennett2016}.
The NASA-WFIRST mission (launch foreseen in 2028) has both microlensing and direct-imaging capabilities.
Through microlensing it is expected to find a total of $\sim$ 1400 bound (to single stars) exoplanets with mass greater than $\sim$ 0.1 M$_\oplus$, including $\sim$200 with mass $\lesssim$ 3 M$_\oplus$\cite{Penny2018}, and 
will vastly increase the number of CBPs detected\cite{Luhn2016}.
We plot in Fig. \ref{fig:MassVSsep} the selection curve for WFIRST.\\
Note that there will be strong complementarity with LISA for circumbinary planets with a separation up to $\sim$ 2 AU.
A joint detection with GWs would be useful to confirm the presence of a CBP by strengthening the lower limit for the planetary mass.

\subsection*{Data availability statement}
Data sharing not applicable to this article as no datasets were generated or analysed during the current study.

\bibliographystyle{aasjournal}
\bibliography{GWExobib}

\leavevmode\thispagestyle{empty}\newpage

\newpage

\end{document}